**Impact of Artificial Intelligence on Economic Theory**

Tshilidzi Marwala

University of Johannesburg

Abstract

Artificial intelligence has impacted many aspects of human life. This paper studies the impact of artificial intelligence on economic theory. In particular we study the impact of artificial intelligence on the theory of bounded rationality, efficient market hypothesis and prospect theory.

**Introduction**

Artificial intelligence is a paradigm where computers or machines are designed to perform tasks that require high level cognition. This is normally achieved by looking at nature and designing machines that are inspired by objects or systems from nature that have been perfected over a long period of time. For example, one can look at how a colony of ants find a shortest distance from its home and the food source and use this to design routing algorithms that are essential for our GPS guides in our cars. The impact of artificial intelligence on major areas of economic sectors is extensive. In the manufacturing industry the application of artificial intelligence to perform tasks that used to be performed by humans will result in extensive job losses. Artificial intelligence has found applications in complex areas in the social, political and economic spaces. For example, Marwala and Lagazio (2011) applied artificial intelligence extensively to model militarized interstate conflict. In this regard the problem of conflict resolution which traditionally required human intuition now involves using computers empowered with artificial intelligence to secure peace. Another application of artificial is the application of artificial intelligence to better design complex structures such as aircrafts. In this regard Marwala et. al. (2015a) and Marwala (2010) were able to use artificial intelligence to create models that are essential to the design of complex systems such as aircraft. In decision making one essential aspect is to be able to secure all information required to make a rational decision. Artificial intelligence has been applied successfully to fill in the gaps that exist in information required to make informed decision. Marwala (2009) applied artificial intelligence to fill in missing information and applied this to decision making in assessing the risks associated with making decisions with incomplete information. Monitoring the conditions of structures such as bridges is essential for securing safe utilization of essential public goods such as bridges. In this regard Marwala (2012) applied artificial intelligence to monitor the conditions of essential mechanical and electrical engineering structures essential in the electricity industry.

**Bounded Rationality**

One aspect of the artificial intelligence is how this technology changes economic theories. Marwala (2013) applied artificial intelligence to model economic and financial instruments such as the stock markets, derivatives and options. How then does artificial intelligence changes economic theory? For example, Economics Nobel Laureate Herbert Simon (1991) observed that on making decisions rationally one does not have the perfect and complete information to make a fully rational decision. Moreover, one does not have the perfect brain to process such information timely and efficiently and the human brain is not consistent and thus decisions made by a human brain are thus inconsistent as they change depending on other factors such as moods swings. Simon termed decision making under such circumstances bounded rationally. With the advent of artificial intelligence one is able to access information that was hidden and thus not accessible, and is able to use such information consistently by the use of artificial intelligence for decision making and is able to increasingly make such decisions more timely and efficiently due to Moore's Law which states that the processing power of machines is always increasing (Moore, 2006). What does this advent of artificial intelligence mean for the theory of bounded rationality? It means that the bounds in Simon's theory of bounded rationality are in effect flexible due to Moore's Law, advanced signal processing and artificial intelligence and more information on this can be studies in the books by Marwala (2014 and 2015b).

**Efficient Market Hypothesis**

Another economic theory which is influenced by the advent of artificial intelligence is the theory of the efficient market hypothesis developed by Nobel Laureate Eugene Fama (1965). This hypothesis states that it is often difficult to beat the markets because the markets are efficient. The problem, is that because the traders in the market are often not perfect and the information they have is imperfect and incomplete the markets are not efficient. Now what happens to the efficient market theory if the traders in the market are not just people but are a combination of people and artificial intelligence infused computer trader? The more artificial intelligence empowered computer traders we have in the markets the more efficient the markets become and therefore the degree at which markets are efficient depends on the amount of artificial intelligent traders we have in the markets.

**Prospect Theory**

Nobel Laureate Daniel Kahneman and Amos Tversky (1979) proposed the prospect theory that states that when people make decisions with the probability of outcomes known they weigh potential losses against potential gains to make such decisions. The impact of this theory on the markets is extensive. However, it rests on the fact that the decision maker is just a person. What happens to this theory if the decision maker is not just a person but a person who is using an artificial intelligent decision machine? How about if the decision maker is wholly an artificial intelligent machine? The applicability of Prospect Theory solely depends on how much artificial intelligent machine is used to make such a decision.

**Conclusion**

Decision making is more and more involving artificial intelligent machine. This paper described how three economic theories are impacted by the application of artificial intelligent machine in the decision making process. It is found that the use of artificial intelligent machine changes the degrees in which the theory of bounded rationality, efficient market hypothesis and prospect theories are applicable.

Reference:

1. Fama, Eugene (1965). "The Behavior of Stock Market Prices". Journal of Business 38: 34–105. doi:10.1086/294743.Kahneman, Daniel; Tversky, Amos (1979). "Prospect Theory: An Analysis of Decision under Risk" (PDF). Econometrica 47 (2): 263. doi:10.2307/1914185. ISSN 0012-9682.
2. Marwala, Tshilidzi, Boulkaibet, Ilyes, and Adhikari Sondipon. (2015a). *Probabilistic Finite Element Model Updating Using Bayesian Statistics: Applications to aeronautical and mechanical engineering*. John Wiley and Sons  ISBN: 978-1-119-15303-0 (in press).
3. Marwala, Tshilidzi (2015b). *Causality, Correlation, and Artificial Intelligence for Rational Decision Making*. Singapore: World Scientific. ISBN 978-9-814-63086-3.

4. Marwala, Tshilidzi (2014). *Artificial Intelligence Techniques for Rational Decision Making*. Heidelberg: Springer. ISBN 978-3-319-11423-1.
5. Marwala, Tshilidzi (2013). *Economic Modeling Using Artificial Intelligence Methods*. Heidelberg: Springer. ISBN 978-1-84996-323-7.
6. Marwala, Tshilidzi (2012). *Condition Monitoring Using Computational Intelligence Methods*. Heidelberg: Springer. ISBN 978-1-4471-2380-4.
7. Marwala, Tshilidzi; Lagazio, Monica (2011). *Militarized Conflict Modeling Using Computational Intelligence*. Heidelberg: Springer. ISBN 978-0-85729-789-1.
8. Marwala, Tshilidzi (2010). *Finite Element Model Updating Using Computational Intelligence Techniques: Applications to Structural Dynamics*. Heidelberg: Springer. ISBN 978-1-84996-322-0.
9. Marwala, Tshilidzi (2009). *Computational Intelligence for Missing Data Imputation, Estimation, and Management: Knowledge Optimization Techniques*. Pennsylvania: IGI Global. ISBN 978-1-60566-336-4.
10. Moore, Gordon (2006). "Chapter 7: Moore's law at 40". In Brock, David. Understanding Moore's Law: Four Decades of Innovation (PDF). Chemical Heritage Foundation. pp. 67–84. ISBN 0-941901-41-6.
11. Simon, Herbert (1991). "Bounded Rationality and Organizational Learning". Organization Science 2 (1): 125–134. doi:10.1287/orsc.2.1.125